\documentclass{jfm}
\usepackage{graphicx}
\usepackage{amsmath,amssymb}
\usepackage{epstopdf, epsfig}

\usepackage{float}
\usepackage{bm}

\usepackage{color}
\usepackage{mathtools}
\def\strutdepth{\dp\strutbox}
\def\nw#1{\strut\vadjust{\kern-\strutdepth\vtop to0pt{\vss\hbox to\hsize
{\hskip\hsize\hskip5pt$\leftarrow$\hss\strut}}}{\em #1}}

\shorttitle{Directional spreading of a viscous droplet on a conical fibre}
\shortauthor{T. S. Chan, F. Yang, A. Carlson}

\title{Directional spreading of a viscous droplet on a conical fibre}

\author{Tak Shing Chan\aff{1} \corresp{\email{taksc@math.uio.no}}, Fan Yang\aff{2}, Andreas Carlson\aff{1} \corresp{\email{acarlson@math.uio.no}}}

\affiliation{\aff{1}Mechanics Division, Department of Mathematics, University of Oslo, Oslo 0316, Norway
	\aff{2}Department of Mechanical and Aerospace Engineering, Princeton University, Princeton, NJ 08544, USA}
\begin{document}
\maketitle

\begin{abstract}
If a droplet is placed on a substrate with a conical shape it spontaneously starts to spread in the direction of a growing fibre radius. We describe this capillary spreading dynamics by developing a lubrication approximation on a cone and by the perturbation method of matched asymptotic expansions. Our results show that the droplet appears to adopt a quasi-static shape and the predictions of the droplet shape and spreading velocity from the two mathematical models are in excellent agreement for a wide range of slip lengths, cone angles and equilibrium contact angles. At the contact line regions, a large pressure gradient is generated by the mismatch between the equilibrium contact angle and the apparent contact angle that maintains the viscous flow. It is the conical shape of the substrate that breaks the front/rear droplet symmetry in terms of the apparent contact angle, which is larger at the thicker part of the cone than that at its thinner part. Consequently, the droplet is predicted to move from the cone tip to its base, consistent with experimental observations. %We compare our results with the prediction of the pressure gradient model proposed by \cite{lorenceau2004}. 
\end{abstract}
\section{Introduction}\label{intro}
A spherical droplet that comes in contact with a solid substrate will change its shape in order to minimize its total surface energy by generating a spreading motion. Droplet spreading on flat substrates has been widely studied and is quite well understood \citep{T79,H83,C86,CW89,BB93,BEIMR09,CAR12}. If the flat substrate has a constant equilibrium contact angle, the center of mass of the droplet will not change its position along the substrate. One way to generate a directional droplet spreading is to manipulate the chemical composition or the micro/nano-scale structure of the substrate to make the equilibrium contact angle vary on the substrate \citep{Brochard1989,Chaudhury1539,Santos1995,Yutaka2005,Moosavi2011,Yan2017}, where the difference in the contact angles at the front and back of the droplet generates its motion. An alternative to surface coatings is to instead change the macroscopic shape of the substrate to a geometric structure with a closed surface where its normal is not the same everywhere e.g. a cone-like structure. A droplet placed on a fibre with the shape of a cone spontaneously starts to move in the direction of a growing cone radius when the flow is dominated by capillary forces \citep{Bico2002,lorenceau2004}. %despite that the droplet size is much smaller than the capillary length $\ll (\rho g/\gamma)^{\frac{1}{2}}$ with $\gamma$ the surface tension coefficient, $g$ gravitational acceleration and $\rho$ the liquid density. 
In fact, the principle of self-propelled droplets by tuning the macrosopic geometry of the substrate has been widely exploited by living creatures, where plants \citep{Lui2015} and animals \citep{Zheng2010,Wang9247,Duprat2012} have evolved thin structures that generate droplet motion. Cacti that reside in arid regions have developed conical spines for water collection from humid air, which also transport the water droplets from the spine tip to its base for adsorption \citep{Lui2015}. \cite{Zheng2010} showed that a similar principle of directional water collection appears on wetted spider silk, where small water droplets condense at the thinner part (the joint) and move to the center of the thicker part (knots) where multiple droplets in time coalescence to create a large droplet. It was recently shown by \cite{Chen2018} that the plant {\em Sarracenia trichome} has developed a solution for droplet transport that generate velocities three orders of magnitude larger than what is found on the spines of cacti by careful design of its macroscopic geometry and its microscopic structure. The biological transport solution in {\em Sarracenia trichome} was mimicked in microchannels as a solutions for rapid droplet transport. Insects, on the other hand, are in general interested in getting rid of the unwanted weight of water droplets. Water striders have legs covered with tilted conical setae, which are elastic and hydrophobic, and aid the removal of water droplets (\cite{Wang9247}). Controlled droplet motion has a broad industrial relevance for controlled chemical reactions, fabrication of materials that can maintain a dry or a wet state or new materials for water harvesting, where recent advances have found inspiration in Nature \citep{Park2016,CHEN2018274,Chen2018}.

When the droplet size, $V^{1/3}$ with $V$ the droplet volume, is smaller than the capillary length $\equiv (\gamma/\rho g)^{\frac{1}{2}}$ with $\gamma$ the surface tension coefficient, $g$ the gravitational acceleration and $\rho$ the liquid density, its  shape and directional movement on a conical fiber is expected to be generated by the capillary forces. \cite{lorenceau2004} studied this droplet dynamics experimentally, where the motion was rationalised by the aid of a theoretical model. The authors first consider the pressure $p_{cy}$ inside an equilibrium barrel-shaped droplet on a cylindrical fiber of radius $R$, which was derived by \cite{CARROLL1976488}, and expressed as  $p_{cy}=\frac{2\gamma}{R+H}+p_o,$
where $H$ is the maximal thickness of the droplet and $p_o$ is the surrounding pressure in the air. It is then assumed that the pressure $p_{co}(x)$ inside a moving droplet on a conical fiber can be written in the same functional form, but with a replacement of $R$ and $H$ by the cone radius $R(x)$ and the droplet thickness $H(x)$ respectively. The substitution of these constants to variables gives rise to a pressure gradient $\frac{dp_{co}}{dx} =-\frac{2\gamma}{\left[ R(x)+H(x)\right]^2}\left(\frac{dR}{dx}+\frac{dH}{dx} \right)$ inside the droplet along the fiber's central axis $x$. Despite that this pressure gradient model has been widely adopted to explain the motion of droplets on conical fibres \citep{Zheng2010,Li2013,Li2016,Chen2018}, to the best of our knowledge, there are no precise measurements that have shown a direct comparison with this theory for the droplet shape and spreading velocity.%Moreover, there has not been any precise measurement of the droplet velocity and shape in order to verify the pressure gradient model.
 
A difference in the equilibrium contact angle at the advancing and the receding contact line of a droplet is known to generate directional spreading of droplets. For a droplet moving on a conical fiber, the equilibrium contact angle is constant, but it still moves and it remains unclear how a difference in the apparent contact angle between the advancing and receding front is generated. We are particularly curious to see how the stresses are distributed inside the droplet and how these compare to those predicted by the pressure gradient model proposed by \cite{lorenceau2004}. To do this, we start by considering a viscous droplet of dynamic viscosity $\eta$ that moves on a fiber when inertia can be neglected, i.e., Reynolds number, $ Re\equiv\frac{\rho UV^{1/3}}{\eta}\ll1$ with the droplet size $V^{1/3}\ll (\gamma/\rho g)^{\frac{1}{2}}$, $U$ the spreading velocity and the motion is dominated by the capillary force but hindered by viscous friction, i.e., the capillary number $Ca\equiv \frac{\eta U}{\gamma}\ll1$. A characteristic feature of slowly spreading  viscous droplets is that it maintains a quasi-static shape during wetting, where viscous stresses are predominantly located in the vicinity of the contact line and balanced by the capillary stress through interface deformations. Together, this makes the problem well suited to be studied by the perturbation method of matched asymptotic expansions, which has been widely used to describe viscous spreading of droplets on flat substrates \citep{T79,Wilson1982,H82,H83,C86,E04b,E05,Len2006,Savva2009,SnE10,CGS11}. We develop here a similar approach, where we consider the viscous spreading of an axisymmetric droplet on a conical fibre by combining the lubrication theory and a perturbation method of asymptotic matching to decipher the physical mechanism that sustains the directional droplet motion.
%In this article, we develop a mathematical model to describe the physical mechanisms investigate the motion of an axisymmetric droplet on a cone.
%A lubrication model for flows on a conical geometry is developed in section \ref{sec:lubrication} and a perturbation method of asymptotic matching is implemented in section \ref{sec:asymptotics}. The shapes and the velocities of the droplet are computed for different droplet positions. In section \ref{sec:results}, we compare the results obtained by the two approaches and discuss the driving mechanism that maintains the motion of the droplet. Our results are compared with the prediction of the pressure gradient model proposed by \cite{lorenceau2004} in section \ref{sec:discussion}.

\begin{figure}
\begin{center}
\includegraphics[width=0.7\textwidth]{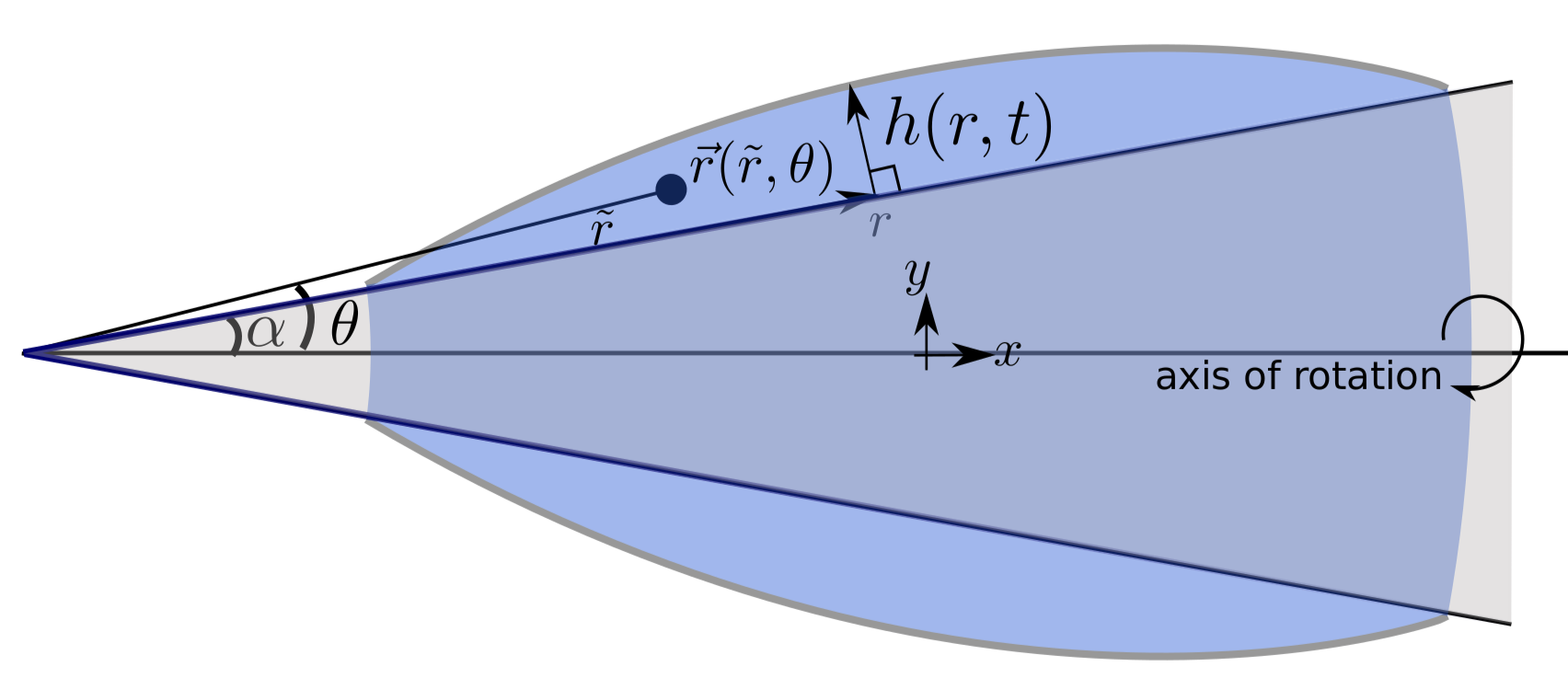}
\caption{Sketch of a droplet on a conical fibre with an angle $\alpha$. The droplet shape is described by the height $h(r,t)$ from the substrate to the free surface as a function of the radial distance from the vertex along the substrate $r$ and time $t$. By using spherical cooridnates, the position vector $\vec{r}$ of a fluid element in the droplet is described as a function of the radial distance from the vertex $\tilde{r}$ and the azimuthal angle $\theta$.}\label{setup}
\end{center}
\end{figure}
 
\section{Mathematical formulation}\label{form}
We consider a droplet in contact with a solid conical fiber with an angle $\alpha$ as shown in figure \ref{setup}. We assume the shape of the droplet is symmetric around the central axis of the cone. The droplet shape is described by the height $h(r,t)$ from the substrate to the free surface as a function of the radial distance from the vertex along the substrate $r$ and time $t$. As the droplet spontaneously starts to move on the cone, an incompressible viscous flow is generated. Thus, the flow inside the droplet is described by Stokes equations 
\begin{eqnarray}\label{stokes1}
\eta \nabla^2 {\bf u}-{\bf \nabla} p=0\, ,
\end{eqnarray}
and the continuity equation reads
\begin{eqnarray}\label{stokes2}
{\bf \nabla} \cdot{\bf u}=0\, ,
\end{eqnarray}
where ${\bf u}$ is the velocity and $p$ is the pressure. Moreover, we only consider small droplets, with a shape unaffected by gravity, i.e. the Bond number $Bo\equiv\rho g V^{2/3}/\gamma\ll1$. We neglect any influence of the air surrounding the droplet as its viscosity is orders of magnitude smaller than the liquid viscosity.

To describe the fluid flow and the droplet motion, eq. (\ref{stokes1}) and (\ref{stokes2}) need to be accompanied by several boundary conditions.
At the free surface, the tangential stress is zero as we neglect the viscous effects in the air. The normal stress $\sigma^{f}_\textrm{n}$ is described by Young-Laplace's law
\begin{equation}\label{ka}
\sigma^{f}_\textrm{n}=\gamma \kappa\, ,
\end{equation}
where $\kappa$ is the curvature of the interface.  

At the solid/fluid boundary, the velocity normal to the substrate is zero and we have a tangential velocity $u^{s}_\textrm{t}$ described by the Navier-slip condition 
\begin{equation}\label{us}
u^{s}_\textrm{t}=\frac{\lambda}{\eta} \sigma^{s}_\textrm{t}\, ,
\end{equation}
where  $ \sigma^{s}_\textrm{t}$ is the shear stress  parallel to the substrate and $\lambda$ is the slip length. Slippage of fluid along the substrate is well known to regularize the viscous stress singularity at the moving contact line where the liquid, air and solid phase intersect. In addition, we need to specify the slope of the free interface at the contact line, given by the equilibrium contact angle $\theta_e$ that follows from Young's law $\cos\theta_e=\frac{\gamma_{SL}-\gamma_{SV}}{\gamma_{\gamma}}$, and is independent of the contact line velocity. $\gamma_{SL}$ is the liquid/solid surface tension coefficient and $\gamma_{SV}$ is the solid/air surface tension coefficient.

\subsection{Lubrication approximation on a cone (LAC)}\label{sec:lubrication}
For a cone angle $\alpha\ll 1$ and an equilibrium contact angle $\theta_e\ll 1$, the flow inside the droplet is primarily in the radial direction and the droplet is fairly flat. By using these approximations, the Stokes equations (\ref{stokes1}) simplifies to the lubrication equations here given in spherical coordinates
\begin{subeqnarray}
	\label{g}
	\frac{\partial p}{\partial \tilde{r}}=\frac{\eta}{\tilde{r}^2\sin \theta}\frac{\partial}{\partial \theta}\left(\sin \theta \frac{\partial u}{\partial \theta}\right),\\
	\frac{\partial p}{\partial \theta}=0,
\end{subeqnarray}
where $u=u(\tilde{r},\theta)$ is the radial velocity. Note that $\tilde{r}$ is the radial coordinate of a liquid element in the droplet. The boundary conditions are
\begin{subeqnarray}\label{bc}
\frac{\partial u}{\partial \theta}=0\quad \mbox{at}\quad \theta=\alpha+\frac{h}{r},\\
\frac{u}{\lambda}=\frac{1}{r}\frac{\partial u}{\partial \theta} \quad \mbox{at} \quad \theta=\alpha, \slabel{bc2}
\end{subeqnarray}
where $r$ is the radial distance from the vertex along the substrate. 
Solving (\ref{g}) with the boundary conditions (\ref{bc}) gives us the velocity
\begin{eqnarray}
u&=&-\frac{r^2}{\eta}\frac{\partial p}{\partial r}\Bigg\{\left[2\ln\left(\cos\frac{\theta}{2}\right)-2\ln\left(\cos\frac{\alpha}{2}\right)\right]+\left[1-
\cos\left(\alpha+\frac{h}{r}\right)\right]\left[\ln\left(\tan \frac{\theta}{2}\right)-\ln\left(\tan \frac{\alpha}{2}\right)\right]\nonumber\\
&&-\frac{\lambda}{r\sin\alpha}\left[\cos\left(\alpha+\frac{h}{r}\right)-\cos\alpha\right]\Bigg\}. \label{u}
\end{eqnarray}
%=\frac{r^2}{4\mu}\frac{\partial p}{\partial r}\Bigg\{\left(\theta^2-\alpha^2\right)-2\left(\alpha+\frac{h}{r}\right)^2\left[\ln\left(\tan \frac{\theta}{2}\right)-\ln\left(\tan \frac{\alpha}{2}\right)\right]\nonumber\\
%+\frac{4b}{r\sin\alpha}\left[1-\frac{1}{2}\left(\alpha+\frac{h}{r}\right)^2-\cos\alpha\right]\Bigg\}.
%where $\nabla_r^2=\frac{1}{r^2}\frac{\partial}{\partial r}\left(r^2\frac{\partial}{\partial r}\right)$. We also used $\int \ln \tan(\theta/2)\mathrm{d}\theta\approx \int \ln (\theta/2) \mathrm{d}\theta=\theta \left[\ln (\theta/2)-1\right]$+constant in deriving (\ref{final}).
The dynamics of the droplet's interface $h=h(r,t)$ is obtained by imposing droplet mass conservation 
\begin{eqnarray}\label{conlub}
\frac{\partial h}{\partial t}+\frac{1}{ r\left(\alpha+ \frac{h}{r}\right)}\frac{\partial}{\partial r}  \int_{\alpha}^{\alpha+h/r}ur^2\sin\theta d\theta=0.
\end{eqnarray}
Note that $\tilde{r}\approx r$ for flat droplets. The pressure inside the liquid is determined by Young-Laplace's equation
%\begin{equation}
%p=-\gamma\kappa, \label{p}
%\end{equation}
(\ref{ka}) with the normal stress $\sigma^{f}_\textrm{n}=-p$.

We next assume that the spreading to be quasi-steady, and the entire droplet moves at a velocity $u_{cl}$ for a small time increment. We expect this assumption to be justified when the capillary number $Ca\equiv\eta u_{cl}/\gamma\ll1$, i.e. the contact line moves slowly as compared to the capillary velocity $\gamma/\eta$. In the frame of the moving droplet, the droplet shape is stationary, i.e. $\partial h/\partial t=0$, and the velocity inside the liquid is $u-u_{cl}$, hence we can reduce the time-dependent lubrication equation (\ref{conlub}) to a stationary form 
\begin{equation}
\int_{\alpha}^{\alpha+h/r}(u-u_{cl})\sin\theta\mathrm{d}\theta=0. \label{steadyThin}
\end{equation}
In addition we have imposed a zero flux condition at the contact line.
To evaluate (\ref{steadyThin}), we substitute (\ref{u}) and (\ref{ka}) with the normal stress $\sigma^{f}_\textrm{n}=-p$ into (\ref{steadyThin}), to obtain
\begin{equation}\label{kap}
\frac{\partial \kappa}{\partial r}=\frac{Ca}{F(h,r,\alpha,\lambda)}, 
\end{equation}
and
\begin{eqnarray}
&&F(h,r,\alpha,\lambda)=\frac{2r^2}{\cos \alpha-\cos\left(\alpha+h/r\right)}\cdot\nonumber\\
&&\left\{\cos^2\frac{\alpha}{2}\left[2\ln(\cos\frac{\alpha}{2})-1\right]-\cos^2\left(\frac{\alpha+\frac{h}{r}}{2}\right)\left[2\ln\left(\cos\left(\frac{\alpha+\frac{h}{r}}{2}\right)\right)-1\right]\right\}\nonumber\\
&&-r^2\left\lbrace 2\ln\left(\cos\frac{\alpha}{2}\right)+\left[1-\cos\left(\alpha+\frac{h}{r}\right)\right]\ln\left(\tan \frac{\alpha}{2}\right)+\frac{\lambda}{r\sin\alpha}\left[\cos\left(\alpha+\frac{h}{r}\right)-\cos\alpha\right]\right\rbrace\nonumber\\
&&+\frac{r^2\left[1-\cos\left(\alpha+h/r\right)\right]}{\cos \alpha-\cos\left(\alpha+h/r\right)}\int_{\alpha}^{\alpha+h/r}\sin\theta\ln\left(\tan \frac{\theta}{2}\right)\mathrm{d}\theta. \label{F}
\end{eqnarray}

The curvature at the free surface can be written as
\begin{equation}\label{dd}
\kappa=\frac{h^{''}}{(1+h^{'2})^{3/2}}-\frac{1}{(r\sin\alpha 
	+h\cos\alpha )\left[1+\left(\frac{h'+\tan\alpha}{1-h'\tan\alpha }\right)^{2}\right]^{1/2}},
\end{equation}
where the first term is the curvatures along the radial and the second term the curvature along the azimuthal directions. We note that curvature along the azimuthal direction is derived by using a rotation matrix using the cone angle $\alpha$ for the axes.
%Now we look at the series expansion of  
The boundary conditions for $h(r)$ at the receding contact line $r=r_{r}$
\begin{subeqnarray}\label{bcr}
h(r=r_{r})&=&0, \\
h^{'}(r=r_{r})&=&\theta_{e}, 
\end{subeqnarray}
and the advancing contact line $r=r_{a}$
\begin{subeqnarray}\label{bca}
h(r=r_{a})&=&0, \\
h^{'}(r=r_{a})&=&-\theta_{e}. 
\end{subeqnarray}
%and 

In the following, all the lengths are rescaled by the $V^{1/3}$ with the volume $V$ given by
\begin{equation}\label{vol}
V=2\pi\int^{r_a}_{r_r}\int^{\alpha+h/r}_{\alpha}r^2\sin\theta d\theta dr,
\end{equation}
but we use the same symbols for all rescaled quantities. Note that (\ref{kap}) has the same form after rescaling and the model parameters that dictate the dynamics are the cone angle $\alpha$, the equilibrium contact angle $\theta_e$ and the slip length $\lambda$. The droplet profile $h(r)$ and the capillary number $Ca$ are determined by solving (\ref{kap}) with the boundary conditions (\ref{bcr}) and (\ref{bca}) by using the shooting method \citep{press07}.

\subsection{Asymptotic analysis (AM)}\label{sec:asymptotics}
We now turn to a description based on the method of matched asymptotic expansions. The droplet size and the slip length differ by several orders of magnitude and we expect that the governing forces are different at these two length scale. In the vicinity of the contact line, denoted as the inner region, the flow is maintained by the balance of capillarity and viscous force. Away from the contact line, the droplet is considered to be quasi-static with a shape only determined by the capillary force, denoted as the outer region. In the following, we summarize our analysis, which builds on the work by \cite{E05}.

\subsubsection{Inner solutions} \label{insol}
In the inner regions, the characteristic length is the slip length, which is assumed to be much smaller than the local radius of curvature of the cone, and allows us to treat the flow as two-dimensional. Since we have an equilibrium contact angle $\theta_e\ll1$ and cone angle $\alpha\ll 1$, the two-dimensional flow can be described by the lubrication equation \citep{Batch67,Oron97} giving us the droplet shape $h(r)$ in the inner region. At the droplet's trailing edge, i.e., the receding inner region, we define the interfacial profile as $h_r=h_r(x_r)$, where $x_r\equiv r-r_r$ is the distance along the substrate from the contact line position, i.e., the profile is determined by the lubrication equation
\begin{equation}
\label{lub1}
h_r'''(x_r) = \frac{3{\rm Ca}}{h_r^2(x_r) + 3\lambda h_r(x_r)}.
\end{equation}
The prime symbol represents the derivative with respect to the independent variable. Equation (\ref{lub1}) is complemented by the boundary conditions at the substrate where the height

\begin{equation}
h_r(x_r=0) = 0,
\end{equation}
and the profile slope is given by the equilibrium angle
\begin{equation}
h_r'(x_r=0) = \theta_e.
\end{equation}
By following the analysis by \cite{E05}, the asymptotic behavior of $h_r$ for $x_r/\lambda\gg 1$ is

\begin{equation}
\label{inpro}
h_r(x_r) =(3Ca)^{1/3}\left[ \frac{\theta_e\kappa_y x_r^2}{6b}  + b_y x_r\right],
\end{equation}
where
\begin{equation}\label{eq:kb}
\kappa_y = \left(\frac{2^{1/6}\beta}{\pi Ai(s_1)}\right)^2, \quad
b_y = \frac{-2^{2/3}Ai'(s_1)}{Ai(s_1)}, \quad \beta^2=\frac{\pi\exp\left[-\theta_e^3/(9Ca)\right]}{2^{2/3}}.
\end{equation}
$Ai$ is the Airy function and $s_1$ needs to be determined from the asymptotic matching. At the advancing droplet edge, $x_a\equiv r_a-r$ is defined as the distance from the contact line along the substrate, and has a profile $h_a=h_a(x_a)$ described by  
\begin{equation}
\label{lub2}
h_a'''(x_a) = -\frac{3{\rm Ca}}{h_a^2(x_a) + 3\lambda h_a(x_a)},
\end{equation}
and complemented with the boundary conditions at the substrate where the height

\begin{equation}
h_a(x_a=0) = 0,
\end{equation}
and the slope is given by the equilibrium angle
\begin{equation}
h_a'(x_a=0) = \theta_e.
\end{equation}

The asymptotic behavior for $x_a/\lambda\gg 1$ is in a functional form of the Cox-Voinov relation \citep{E05}
\begin{equation}
\label{inadv}
h'_a(x_a)^3 =\theta_e^3+9Ca\ln(e\theta_e x_a/3\lambda),
\end{equation}
where $e=2.7182...$ is the Euler's number.

\subsubsection{Outer solution} \label{outsol}
At the length scale of the droplet size, i.e., the outer region, the dominant force is capillarity. Viscous effects appear only as a small correction in the higher order terms of $Ca$ as we have $Ca\ll1$.
We define the outer solution as $\bar{h}(r)$ and expand it in series of $Ca$
\begin{equation}
\bar{h}(r) =\bar{h}_0(r)+Ca\bar{h}_1(r)+\mathcal{O}(Ca^2).
\end{equation}
We solve for the leading order term $\bar{h}_0(r)$ from the condition of constant interfacial curvature $\kappa$, which can be expressed by the relation  
%\begin{equation}\label{kafiber}
%\kappa=\frac{h_{\rm out}^{''}}{(1+h_{\rm out}^{'2})^{3/2}}-\frac{1}{(r_0+h_{\rm out})(1+h_{\rm out}^{'2})^{1/2}}=c,
%\end{equation}

\begin{equation}\label{kafiber}
\kappa=\frac{\bar{h}_0^{''}(r)}{(1+\bar{h}_0^{'2}(r))^{3/2}}-\frac{1}{(r\sin\alpha 
 +\bar{h}_0(r)\cos\alpha )\left[1+\left(\frac{\bar{h}_0'(r)+\tan\alpha}{1-\bar{h}_0'(r)\tan\alpha }\right)^{2}\right]^{1/2}}.
\end{equation}
Note that this expression is the same as (\ref{dd}) with a replacement of $h(r)$ by $\bar{h}_0(r)$.
The value of the curvature $\kappa$ is determined together with the relation between the volume $V$ and the profile $\bar{h}_0$ of the droplet, that is
\begin{equation}\label{volume}
V=2\pi\int^{r_a}_{r_r}\int^{\alpha+\bar{h}_0/r}_{\alpha}r^2\sin\theta d\theta dr=1.
\end{equation}
%Note that all the lengths are rescaled by $V^{1/3}$ with $V$ the dimensional droplet volume.

The contact angle of the static profile $\bar{h}_0(r)$ with the substrate at the receding part is used to define the receding apparent contact angle $\theta_{r}$, while the advancing droplet edge is defined by the advancing apparent contact angle $\theta_{a}$. Two  boundary conditions are required to solve the second order ordinary differential equation (\ref{kafiber}), where we impose the following conditions at the receding contact line
\begin{eqnarray}
\bar{h}_0(r=r_{r})&=&0, \\
\bar{h}_0'(r=r_{r})&=&\theta_{r}. 
\end{eqnarray}
The position of the advancing contact line and the advancing apparent contact angle are given by the conditions

\begin{eqnarray}
\bar{h}_0(r=r_{a})&=&0, \\
\bar{h}_0'(r=r_{a})&=&-\theta_{a}. 
\end{eqnarray}

To match to the inner solution at the receding contact line region, only the leading order term $\bar{h}_0(r)$, i.e. the static profile, is required.
The asymptotic behavior of $\bar{h}_0(r)$ near the contact line is determined by a Taylor expansion,

\begin{equation}\label{eq:out}
\bar{h}_0(r) = \theta_{r} (r-r_r) + \frac{1}{2} \kappa_{r} (r-r_r)^2 + \mathcal{O}\left((r-r_r)^3 \right).
\end{equation}

At the advancing droplet edge, the first order correction $\bar{h}_1(r)$ is required in order to match to the logarithmic behavior of the inner solution in (\ref{inadv}) \citep{E05}. However, there is no  analytical solution for the correction term for a cone geometry. Instead of computing the exact expression as done by \cite{E05} for a flat substrate, we assume that the outer solution near the advancing contact line has the functional form  \begin{equation}\label{asy_outa}
\bar{h}'^3 =\theta_a^3+9Ca\ln\left[ c_a(r_a-r)\right], 
\end{equation}
where $c_a$ is an adjustable parameter but appears in the logarithm and has only a weak effect on the results.

\subsubsection{Matching}
We are now in a position to perform the matching between the asymptotic behavior of the inner and outer solutions to determine the unknown quantities; $Ca$, $s_1$ and $\theta_r$. To do this, we see that three matching conditions are required. By comparing the asymptotic behavior of the inner solution (\ref{inpro}) and the outer solution (\ref{eq:out}) at the receding region, we find the matching conditions

\begin{equation}\label{sfcondition1}
\theta_{r}=(3Ca)^{1/3}b_y,
\end{equation}

\begin{equation}\label{sfcondition2}
\kappa_{r}=\frac{(3Ca)^{1/3}\kappa_y\theta_e}{3\lambda}.
\end{equation}

At the advancing contact line region, we match the inner solution (\ref{inadv}) to the outer solution (\ref{asy_outa}), the advancing apparent contact angle becomes
\begin{equation}
\label{sfcondition3}
\theta_a^3 =\theta_e^3+9Ca\ln(c/\lambda),
\end{equation}
with $c\equiv{e\theta_e /3c_a}$ that is treated as an adjustable parameter, which we fix here $c=0.04$ by fitting the results of asymptotic matching and lubrication approximation (see section \ref{sec:results}). Since $c$ appears inside the logarithm, it has only a weak influence on the results.

\subsubsection{Completely wetting substrate, $\theta_e=0^{\circ}$}
In the case of a completely wetting droplet, a Landau-Levich-Derjaguin film \citep{LL42,DeGennesWSD} will be deposited on the substrate at the receding contact line as the droplet moves. We can assume that the receding apparent contact angle $\theta_r$ is zero for any value of $Ca$ and matching at the receding region is not needed. The advancing apparent contact angle $\theta_a$ is obtained by solving for the static outer solution governed by (\ref{kafiber}) with $\theta_r=0$. Once $\theta_a$ is computed, the capillary number $Ca$ is determined  by the condition (\ref{sfcondition3}) at the advancing contact line
\begin{equation}
\label{comca}
Ca=\frac{\theta_a^3 }{9\ln(c/\lambda)}.
\end{equation}

\section{Results} \label{sec:results}

\begin{figure}
\begin{center}
\includegraphics[width=1.0 \textwidth]{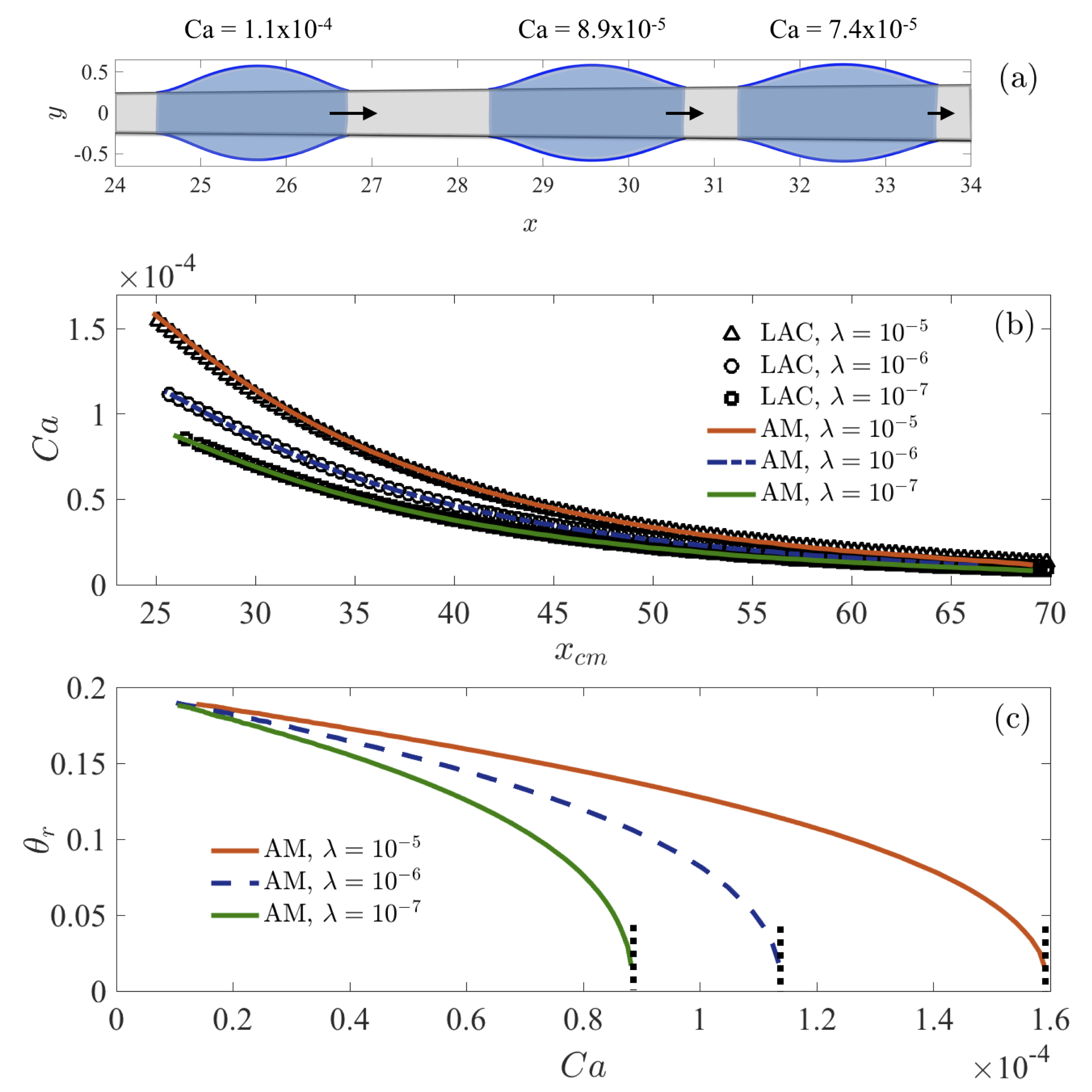}
\caption{(a) The blue curves are the droplet shapes predicted by the lubrication approximation on a cone (LAC) for a slip length $\lambda=10^{-6}$ and $\alpha=0.01$. The grey area represents the conical fiber. $x$ and $y$ are the coordinates respectively along and perpendicular to the axis of rotation. The equilibrium contact angle is $\theta_e=0.2$ rad ($11.5^{\circ}$) and the cone angle is $\alpha=0.01$ rad ($0.57^{\circ}$). similar to the experiments by \cite{lorenceau2004}. (b) The droplet capillary number shown as a function of the droplet's center of mass $x_{cm}$ for three different slip lengths using the  LAC and the asymptotic matching (AM). (c) The receding apparent contact angle $\theta_r$ is plotted as a function of $Ca$ computed by asymptotic matching. The vertical dotted lines indicate the critical capillary numbers $Ca_c$.}\label{fig2}
\end{center}
\end{figure}

%\begin{figure}
%\begin{center}
%\includegraphics[width=0.7\textwidth]{newnewfig2.eps}
%\caption{Test !!}\label{fig2}
%\end{center}
%\end{figure}

Next, we solve the two mathematical models, the lubrication approximation on a cone (LAC) and the asymptotic matching (AM), to predict the directional spreading dynamics of droplets on a conical fibre. There are several important physical parameters that can influence the spreading phenomenon, in particular the cone angle $\alpha$, the slip length $\lambda$ and the equilibrium contact angle $\theta_e$ and we show how these parameters influence the dynamics i.e., the droplet's capillary number $Ca$ and shape. The LAC given by (\ref{kap}-\ref{dd}) is solved by using the shooting method with boundary conditions (\ref{bcr},\ref{bca}). The solution for the AM is obtained by solving the matching conditions (\ref{sfcondition1}-\ref{sfcondition3}) together with the constant curvature condition (\ref{kafiber}) for the static outer solution, which allows us to determine $Ca$, $s_1$, $\theta_r$ and $\theta_a$. The droplet center of mass $x_{cm}$ is used to quantify its position on the cone and given by,  
\begin{equation}\label{xcm}
x_{cm}=2\pi\int^{r_a}_{r_r}\int^{\alpha+h/r}_{\alpha}r^3\sin\theta d\theta dr.
\end{equation}

\subsection{Droplet velocity on a cone}
The droplet shape during directional spreading is shown in figure \ref{fig2}a, where the droplet moves from left to right for a cone with $R=\alpha x$ i.e. the droplet moves to the thicker part of the cone but decelerates along the way, qualitatively consistent with experimental measurements \citep{lorenceau2004}. We start by determining the droplet velocity along the cone, illustrated by the capillary number $Ca$ as a function of $x_{cm}$ and shown for different slip lengths in figure \ref{fig2}(b). The predictions from the two mathematical models (LAC and AM) are in excellent agreement although the slip lengths are varied two orders of magnitude. We see that the droplet velocity is not a linear function of the cone radius $R$ and a critical capillary number $Ca_c$ where no solution is predicted by the LAC or the AM. To understand why no solution is found, we show how the receding apparent contact angle $\theta_r$ as predicted by asymptotic matching as a function of $Ca$ in figure \ref{fig2}(c). The receding contact angle approaches zero as the capillary number gets closer to the critical value $Ca\rightarrow Ca_c$. A vanishing receding contact angle implies that a film is formed at the droplet trail \citep{C86,E04b,SZAFE08}. Also the slip length $\lambda$ can influence the spreading, but its influence on the results is weaker than the other parameters in the system and primarily influence the results for larger $Ca$, in concordance with other moving contact line models \citep{C86,E05,Snoeijer13} as $\lambda$ appears in the logarithmic term in (\ref{sfcondition3}).

\begin{figure}
\begin{center}
\includegraphics[width=0.9\textwidth]{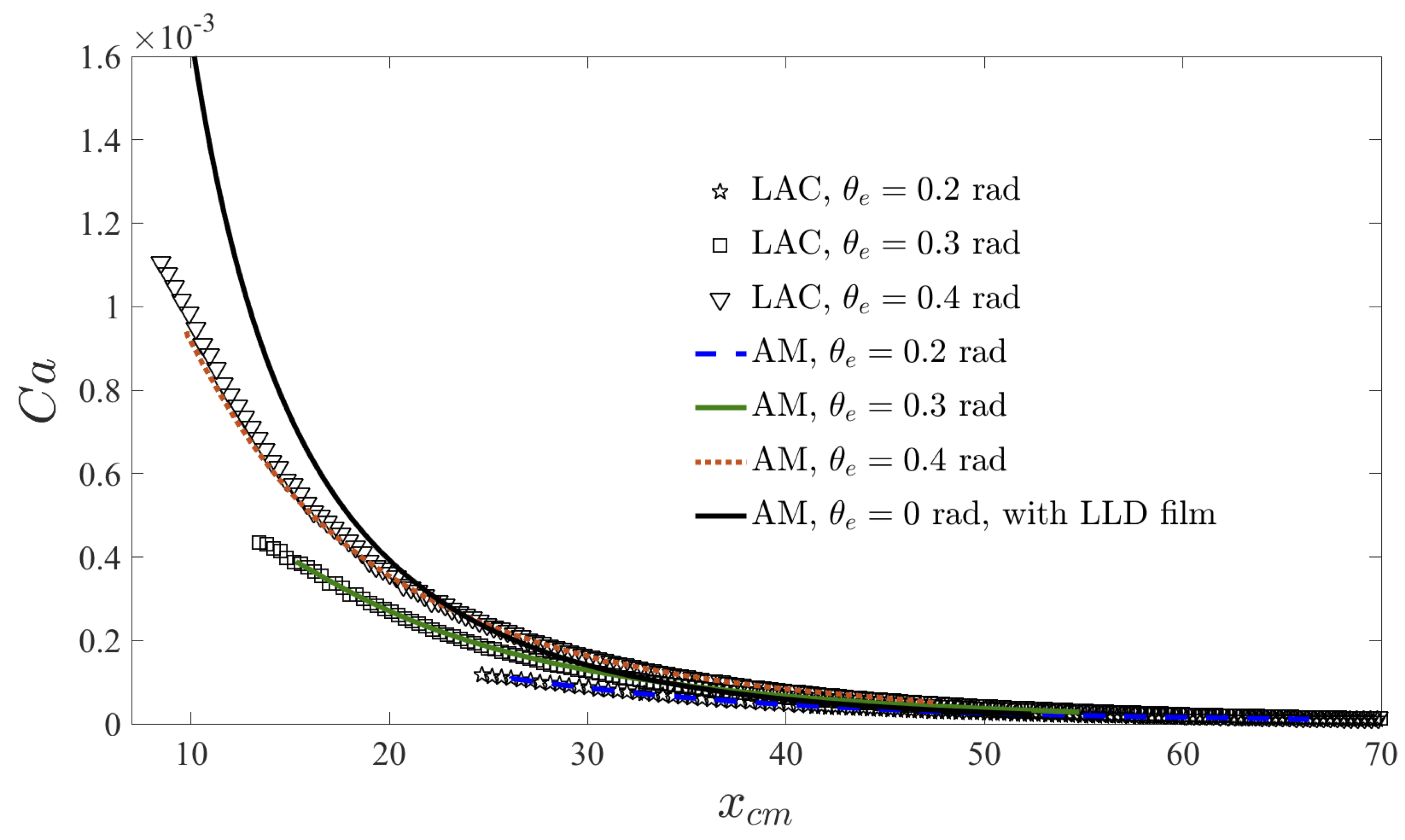}
\caption{The capillary number plotted $Ca$ as a function of the droplet's center of mass $x_{cm}$ for four different equilibrium contact angles $\theta_e$, predicted by the lubrication approximation on a cone (LAC) and by the asymptotic matching (AM). The slip length $\lambda=10^{-6}$ and the cone angle $\alpha=0.01$ rad ($0.57^{\circ}$). Note that the completely wetting $\theta_e=0$ rad is a special case as a Landau-Levich-Derjaguin (LLD) film is deposited on the substrate surface.}\label{ca_theta}
\end{center}
\end{figure}

The droplet wettability i.e the equilibrium contact angle $\theta_e$ also influence the spreading dynamics and we solve the droplet velocity for different $\theta_e$, see figure \ref{ca_theta}. Somewhat counterintuitively a partially wetting droplet at the same position on the cone, but with a larger $\theta_e$ will move faster along the cone. The completely wetting droplet $\theta_e=0$ rad is a special case in figure \ref{ca_theta}, with $\theta_r=0$ at any position $x_{cm}$ on the cone, as a Landau-Levich-Derjaguin film formed at the receding contact line \citep{DeGennesWSD}. We can then compare a completely wetting droplet with a partially wetting droplet at $\theta_r=0$ and at the same  position on the conical fibre. Here, the macroscopic droplet shapes would be the same, as well as their advancing apparent contact angle, but the droplet velocity is determined by (\ref{sfcondition3}) explaining why a completely wetting droplet moves faster than a partially wetting droplet. 

\begin{figure}
\begin{center}
\includegraphics[width=0.8\textwidth]{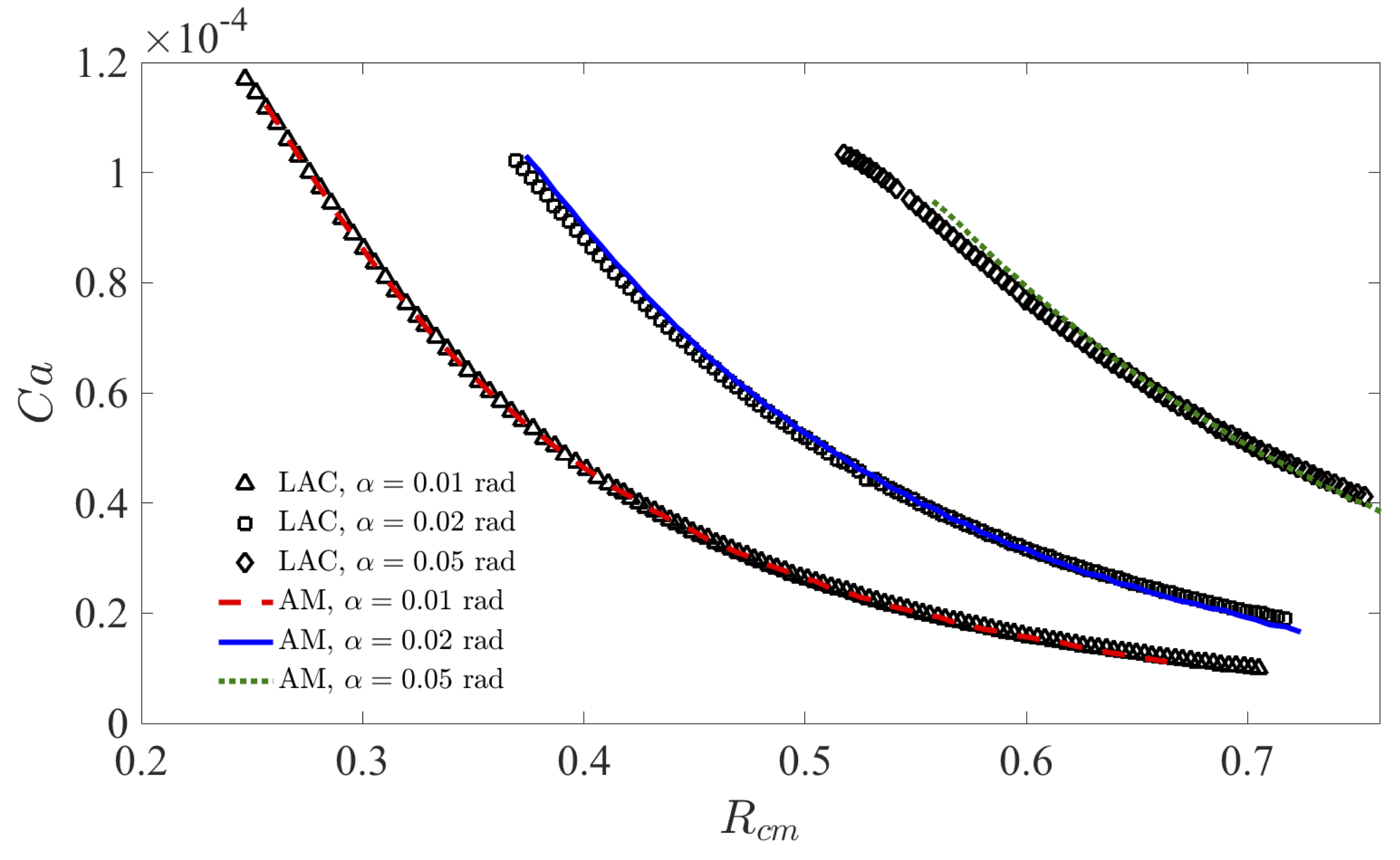}
\caption{The capillary number $Ca$ as a function of the cone radius $R_{cm}$ evaluated for droplet's the center of mass for three different cone angles $\alpha$ by using the approaches of lubrication approximation on a cone (LAC) and the asymptotic matching (AM). The slip length $\lambda=10^{-6}$ and the equilibrium contact angle $\theta_e=0.2$ rad.}\label{Ca_alp}
\end{center}
\end{figure}

To illustrate the dependence on the slope of the cone by varying $\alpha$, we first define $R_{cm}$ as the cone radius at the droplet's center of mass,
\begin{equation}\label{rcm}
R_{cm}=x_{cm}\tan\alpha.
\end{equation} 
Eq. (\ref{rcm}) allows us to compare Ca for different $\alpha$ at the same cone thickness $R_{cm}$, as shown in figure \ref{Ca_alp}. As expected, the droplet moves faster when $\alpha$ is larger for the same $R_{cm}$. Our results highlight how the droplet velocity can be tuned by the control of the macroscopic geometry of the conical substrate shape.     

\subsection{Droplet shape with a mismatch between the apparent and equilibrium contact angle}
\begin{figure}
\begin{center}
\includegraphics[width=0.9\textwidth]{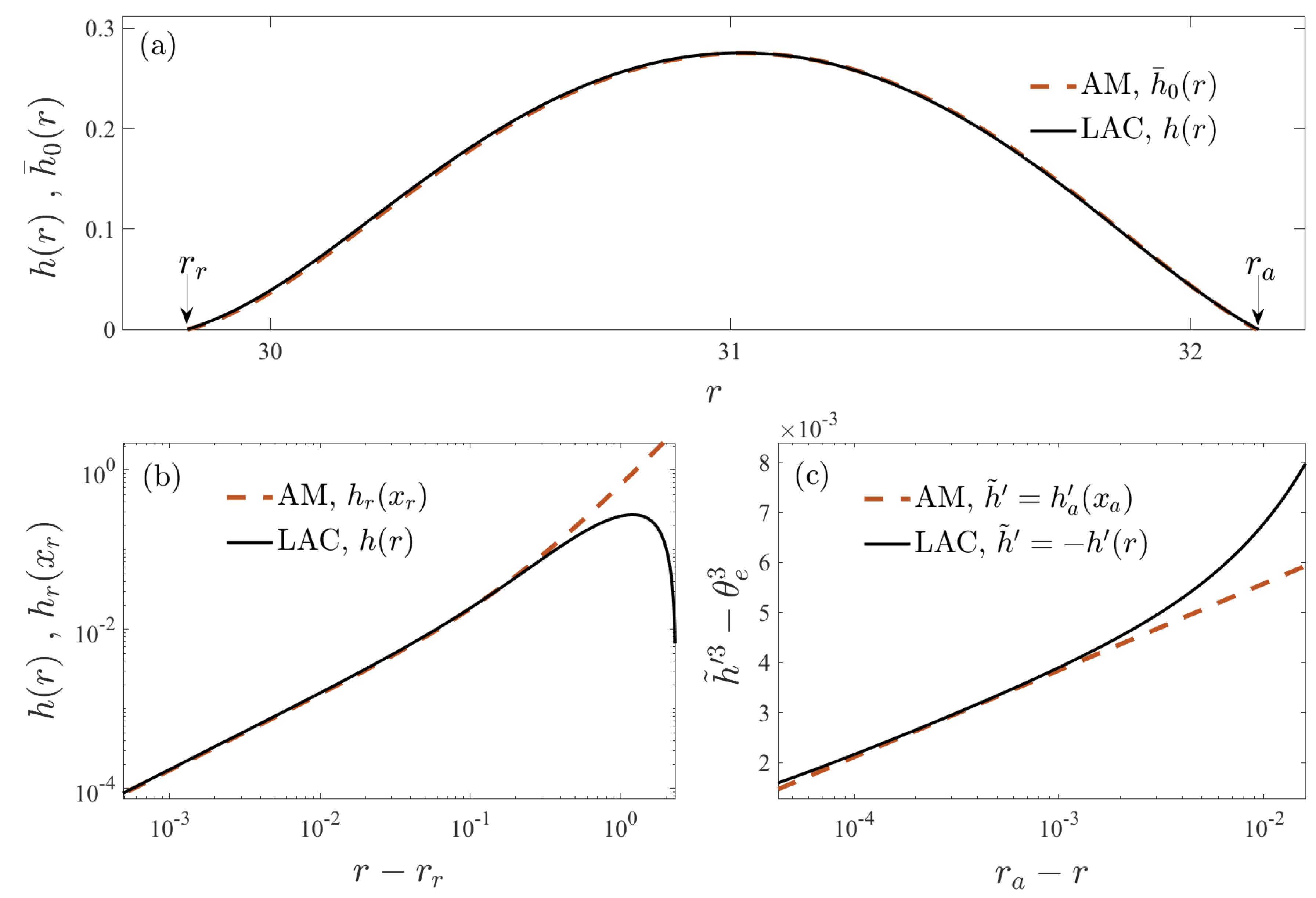}
\caption{(a) The droplet shape predicted by the lubrication approximation on a cone (LAC) $h(r)$ and the asymptotic matching (AM) $\bar{h}_0(r)$ for a slip length $\lambda=10^{-6}$, the equilibrium contact angle $\theta_e=0.2$ rad and the cone angle $\alpha=0.01$ rad. The center of mass of the droplet is $x_{cm}=30.0$ and the capillary number Ca $=8.1$ x $10^{-5}$. $r_r$ and $r_a$ are respectively the receding and advancing contact line positions. (b) The droplet shape $h(r)$ plotted as a function of distance from the receding contact line in log-log scale. The curve for AM is the inner solution $h_r(x_r)$ from (\ref{inpro}). (c) Comparison between LAC and AM in the advancing contact line region. The curve for AM is from (\ref{inadv}), and $\tilde{h}'=-h'(r)$ for the LAC, and $\tilde{h}'=h_a'(x_a)$ for the AM.}\label{profile}
\end{center}
\end{figure}

\begin{figure}
\begin{center}
\includegraphics[width=0.8\textwidth]{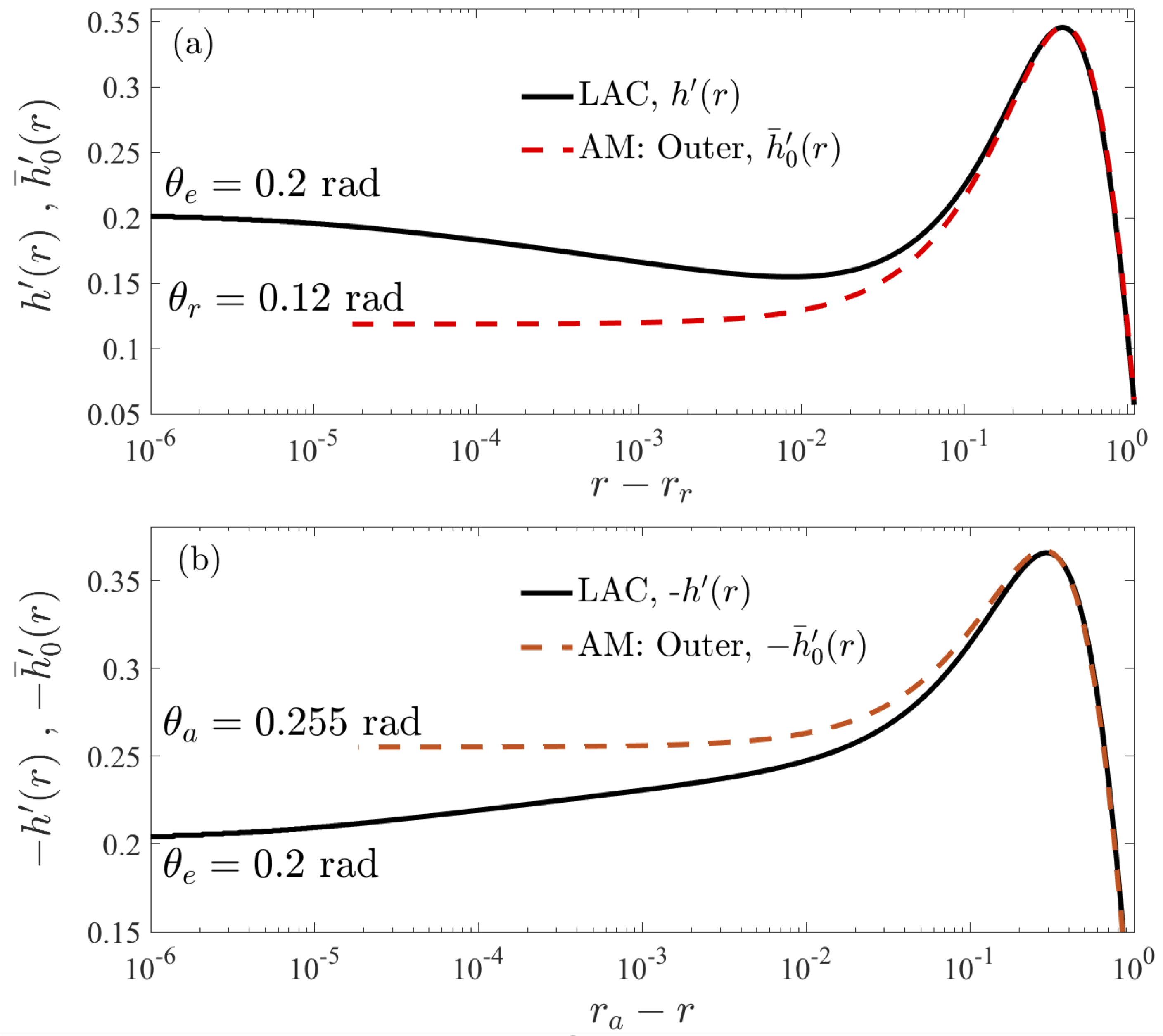}
\caption{(a) The interface slope ($h'$ for LAC and $\bar{h}'_0$ for AM) plotted as a function of the distance from the receding contact line position $r-r_r$ in logarithmic axis.  The slip length $\lambda=10^{-6}$, the equilibrium contact angle $\theta_e=0.2$ rad and the cone angle $\alpha=0.01$ rad. The center of mass of the droplet is $x_{cm}=30.0$ and the capillary number Ca $=8.1$ x $10^{-5}$. LAC: solution obtained by lubrication approximation on a cone. AM: the outer solution obtained by asymptotic matching. (b) The predicted slopes from the LAC and the AM are plotted as a function of the distance from the advancing contact line.}\label{mismatch}
\end{center}
\end{figure}

Since the motion of the droplet is generated by the capillary force, which depends on the interfacial curvature, we would like to know how the droplet shape maintains the viscous flow. In figure \ref{profile}(a), the interface shape is plotted at a scale similar to the droplet size. We see that the prediction from the two theoretical approaches (LAC and AM) are in excellent agreement, illustrating that the droplet shape is quasi-static. In the contact line regions, it is expected that the viscous stress is balanced by the capillary stress. We show the droplet shape $h(r)$ and the difference in slopes by evaluating $\tilde{h}'^3(r)-\theta_e^3$ in the receding and the advancing contact line regions as a function of the distance from the contact line positions, see figure \ref{profile} (b) and (c). In the advancing contact line region, we compare the results from the LCA and the AM in the form of Cox-Voinov relation \citep{V76,C86,E05}. The good agreement between LAC and AM suggests that the force balance assumption in the contact line regions is correct. The interface deforms significantly near the contact line that generate a large capillary stress to maintain the flow inside the two contact line regions. This large interfacial deformation is clearly illustrated in figure \ref{mismatch}, where the local slope of the interface is plotted as a function of the distance from the contact line. As we can see from the curve computed by the LAC in figure \ref{mismatch}(a), the local interface slope decreases from the equilibrium value ($\theta_e=0.2$ rad) at the receding contact line position to a local minimum within a very small distance. The variation of local angle is due to the difference between the equilibrium contact angle and the receding apparent contact angle ($\theta_r=0.12$ rad) determined from the macroscopic shape, see figure \ref{mismatch}(a). In the advancing region, since the apparent contact angle is larger than the equilibrium angle, the interface slope increases with the distance from the contact line position, see figure \ref{mismatch} (b).  The flow and the droplet motion is generated by the mismatch of the equilibrium and apparent contact angles at the receding and the advancing contact line. Because of the asymmetry of the conical shape, the apparent contact angle at the thicker part of the cone is larger than that at the thinner part and the droplet moves spontaneously from the thinner part to the thicker part.

\section{Discussion and conclusion} \label{sec:discussion}
\begin{figure}
\begin{center}
\includegraphics[width=0.8\textwidth]{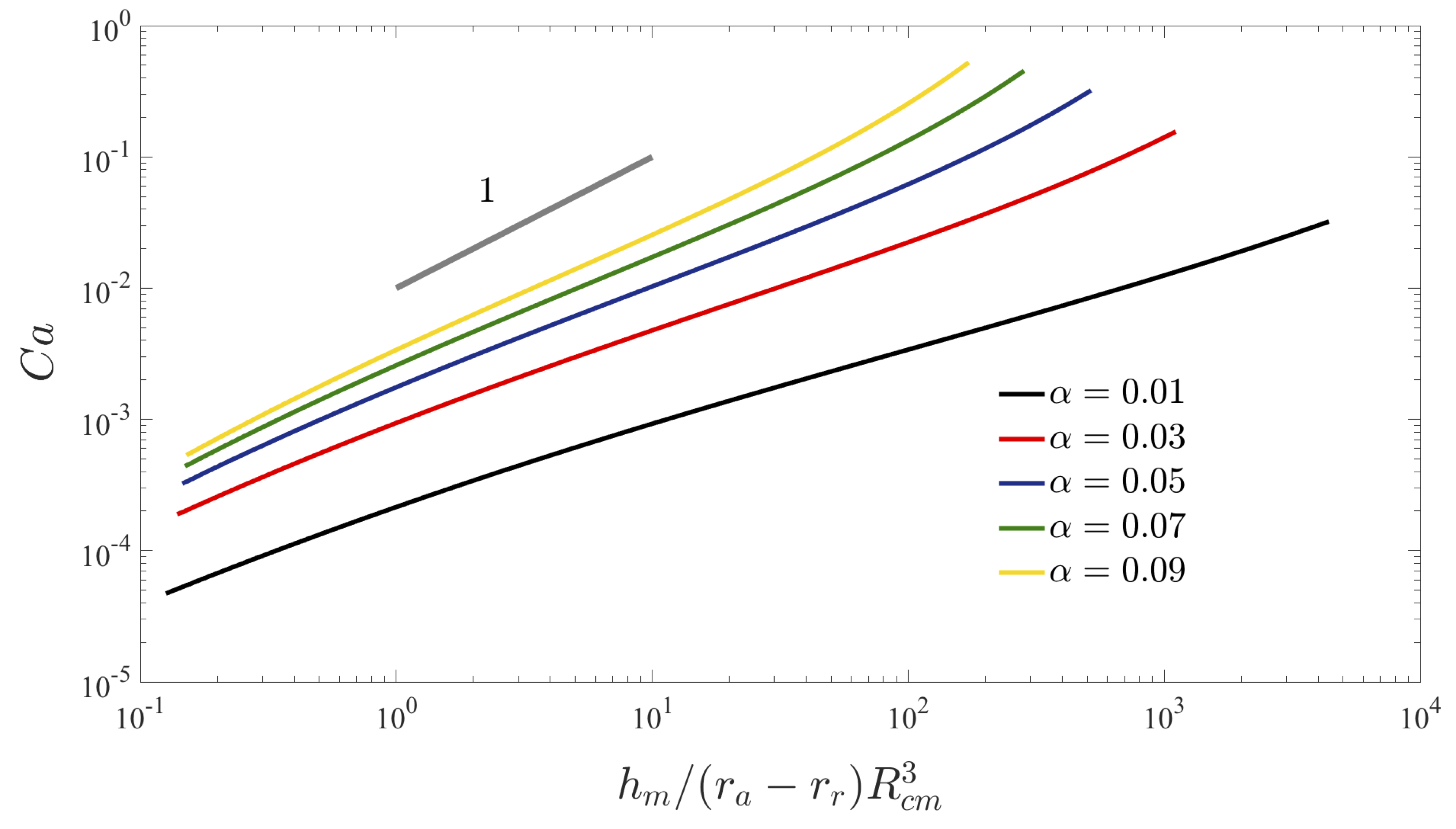}
\caption{The capillary number $Ca$ is computed by using asymptotic matching with (\ref{comca}) for a completely wetting droplet as a function of the quantity $h_m/(r_a-r_r)R_{cm}^3$ (\ref{DQ}). The slip length is $\lambda=10^{-6}$. The straight line has a linear slope.} \label{Ca_geo}
\end{center}
\end{figure}

We have shown that the motion of a fairly flat and viscous droplet on a conical fibre is driven by the difference between the equilibrium and the apparent contact angle, which is not the same at the advancing and the receding front. Moreover, our analysis show that the capillary pressure gradient at the scale of the droplet size is very small and scales with the capillary number (\ref{kap}), which predicts a quasi-static droplet shape. Instead large pressure gradients and strong interfacial deformations are found in the vicinity of the contact line, along with the dominant part of the viscous stresses. Our findings contrast the model proposed by \cite{lorenceau2004} that is based on a capillary pressure gradient at the drop scale to generate the droplet motion, which for complete wetting is described by the relation  
\begin{equation}\label{DQ}
Ca\sim \frac{h_m}{(r_a-r_r)R_{cm}^3},
\end{equation}
where $h_m$ is here the maximum height of the droplet. We plot our results for the completely wetting case $\theta_e=0$ rad and compare the results with the prediction (\ref{DQ}) for several different cone angles $\alpha$ in figure \ref{Ca_geo}. The first observation we make, is that our predictions does not follow the relation (\ref{DQ}). However, our results point to the possibility that if measurements are made over a limited range of $Ca$ or $\frac{h_m}{(r_a-r_r)R_{cm}^3}$ the interpretation of the data can easily be misinterpreted to follow (\ref{DQ}) in particular for larger cone angles.

A mismatch between the equilibrium contact angle and the apparent contact angle is found to dominate the flow and since this difference is not the same at the receding and advancing front a directional droplet spreading motion is generated. Although the equilibrium contact angle is constant on the conical fibre, the apparent angle changes due to its macroscopic asymmetric shape. Our model provides a general description of droplet motion on a conical substrate and can help to better understand droplet motion on slender geometries with varying shapes, which are found in an abundance of engineering systems and in Nature.

\section*{Acknowledgment}
T.S.C. and A.C. gratefully acknowledge financial support from the UiO: Life Science initiative at the University of Oslo. F.Y. is grateful for support from the National Science Foundation (DMS-1514606). A.C. grateful for the financial support from the Norwegian Research Council (263056). 
\bibliographystyle{jfm}
\bibliography{new_all_ref}

\end{document}